\documentclass{ws-ijmpd}
\usepackage[super,compress]{cite}
\usepackage{graphicx}

\begin{document}

\markboth{Angela Di Virgilio et al.}
{Observational and Experimental Gravity}

%
\catchline{}{}{}{}{}
%

\title{Observational and Experimental Gravity  }

\author{Angela Di Virgilio}
\address{INFN-Sezione di Pisa, Largo B. Pontecorvo 3, 56124 Pisa
Italy, 
angela.divirgilio@pi.infn.it}
\author{Wei-Tou Ni}
\address{National Tsing Hua University, Hsinchu, Taiwan, 30013 ROC,  weitou@gmail.com}
\author{Sperello di Serego Alighieri}
\address{INAF - Osservatorio Astrofisico di Arcetri, 50125 Firenze, Italy}
\author{Hung-Yi Pu}
\address{Institute of Astronomy \& Astrophysics, Academia Sinica, Taipei, Taiwan, 10617 ROC}
\author{Sheau-shi Pan}
\address{ Center for Measurement Standards, ITRI, Hsinchu, Taiwan, 30011 ROC}
\maketitle

\begin{history}
\received{Day Month Year}
\revised{Day Month Year}
\end{history}

\begin{abstract} We indicate the progress of experimental gravity, present an outlook in this field, and summarise the Observational/Experimental Parallel Session together with a related plenary talk on gravitational waves of the 2\textsuperscript{nd} LeCosPA Symposium. 
\end{abstract}

\ccode{PACS numbers: 04.80.Cc, 04.80.Nn, 04.80.-y, 95.55.Ym, 98.80.Es}


\section{Progress and Outlook}
For one hundred years since the advent of General Relativity (GR), the first thing in observational and experimental gravity is the observation of light deflection during the solar eclipse in 1919. Since then, the 3 classical tests of GR (perihelion advance, gravitational redshift and light deflection) have been verified\cite{uno} to $10^{-3}-10^{-4}$. With the advent of space age in 1957 and the development of space radio communication, Shapiro proposed a fourth test (Shapiro time delay) of GR that electromagnetic wave packets passing near the Sun would be retarded due to gravity curving of spacetime \cite{due}. Shapiro time delay is measured to agree with GR in terms of Eddington parameter $\gamma$ (equal to 1 for GR) as $1.000021 {\pm}0.000023$ from Cassini spacecraft Doppler tracking\cite{tre}. Lense-Thirring frame dragging has been measured to about 10 percent to agree with GR by satellite laser ranging (SLR) of LAGEOS 1 and LAGEOS 2, and by GP-B gyro relativity experiment\cite{quattro,cinque}. GP-B experiment has also verified the equivalence principle for rotating body to an ultimate precision \cite{sei}.

At present, Lunar Laser Ranging (LLR) test and solar-system radio tests of GR have about the same accuracy. In the near future, interplanetary laser ranging and spacecraft laser ranging in the solar system will improve the accuracy of tests of relativistic gravity by another 3-4 orders of magnitude reaching the second post-Newtonian order\cite{uno}. The observation of precise timing of pulses from pulsars is now catching up the solar system observations and will also reach the second post-Newtonian order soon if not earlier\cite{sette}.
    The development of ring laser gyroscopes makes it possible to measure the absolute angular velocity and to estimate the rotation rates relative to the local inertial frame with ultra-precision. This is ideal for measuring the Lense-Thirring frame dragging on Earth. GINGER (Gyroscopes IN GEneral Relativity) is proposed to measure this frame dragging to 1 percent; GINGERino is under implementation (Sec. 2)\cite{otto}. Technology based on the success of GINGER experiment could be applied to improve the tie between the astronomic reference frame and the solar-system dynamical frame significantly.

Einstein Equivalence Principle is an important cornerstone of GR and metric theories of gravity. From the nonbirefringence of cosmic 
propagation of electromagnetic wave packet, the constitutive tensor of spacetime for local linear gravity-coupling to electromagnetism must be
 of core metric form with an axion (pseudoscalar) degree of freedom and a dilaton (scalar) degree of freedom; from observations it is empirically verified to $10^{-38}$, i.e. to $10^{-4} \times({\it{M}}_{\text{Higgs}}/{\it{M}}_{\text{Planck}})^2$\cite{nove}. This is significant in constraining the infrared behaviour
  of quantum gravity. Empirically, axion is constrained by the non-observation of cosmic polarization rotation (Sec. 2); dilation is constrained by 
  the agreement of CMB spectrum to Planck spectrum\cite{nove,dieci}.

Galileo equivalence principle is experimentally verified by E\"otv\"os-type experiments and free-fall experiments\cite{undici,dodici} to about$10^{-13} $.
 In the next mission/mission proposals for testing Galileo equivalence principle, the improvements are aimed at $2-5$ orders of magnitude\cite{tredici,quattordici}.
    In a series of papers in the 1970s, Rubin, Ford, and Thonnard measured the rotation curves of a number of disk galaxies and found that 
rotation speeds were larger than would be expected from the gravitational attraction arising from the visible mass distribution\cite{quindici,sedici,diciassette}. Authors interpreted their findings as evidence for a new dark matter component. Logically this conflict, known as the missing mass 
problem, could arise from a mass discrepancy, an acceleration discrepancy, or possibly even both. In 1983, Milgrom proposed the 
phenomenological modified Newtonian dynamics (MOND) law for small accelerations \cite{diciotto}. Under this hypothesis, the gravitational
 dynamics become modified when the acceleration is smaller than $a_{0} \sim 10^{-10}$ m s$^{-2}$. However, great efforts in finding the missing 
mass have not been fruitful. Neither the construction of a viable relativistic gravitational theory incorporating the MOND law is fully successful.
 It remains an open question. It is also interesting to note that $a_{0} \sim \Lambda^{1/2}$\cite{diciannove}.

    On the cosmological scale, the discovery of cosmic acceleration have indicated the existence of cosmological constant or the cosmological-constant-like dark energy. The search for a microscopic theory behind this phenomenon may give clue to the microscopic origin of gravity\cite{diciannove}.
    The observation on cosmic structure has confirmed the inflationary scenario that it could have originated from quantum fluctuations in the inflationary period\cite{venti,ventuno}.
    GWs from inflationary period could give imprints of tensor anisotropies and production of B-modes on CMB. CMB polarization observations have constrained tensor to scalar ratios {\it{r}} of inflationary/primordial GWs to be less than $0.07-0.1$\cite{ventidue}. 

Most of the experimental gravitation community are working on GW experiments/observations on various frequency bands from aHz to THz\cite{wei0}. Real-time detection is possible above 300 pHz, while below 300 pHz the detection is possible on GW imprints or indirectly. Advanced LIGO has achieved 3.5 fold better sensitivities with a reach to neutron star binary merging event at 70 Mpc and began its first observing run (O1) in the middle of September 2015 searching for GWs. On September 14 the first GW observation has taken place, and the merger of two Black Holes have been recorded\cite{GW}. We will see a global network of second generation km-size interferometers for GW detection soon. Another avenue for real-time direct detection is from the PTAs. The PTA bound on stochastic GW background already excludes most theoretical models; this may mean we could detect very low frequency GWs anytime too with a longer time scale. Although the prospect of a launch of space GW is only expected in about 20 years, the detection in the low frequency band may have the largest signal to noise ratios. This will enable the detailed study of black hole co-evolution with galaxies and with the dark energy issue. We will see improvement of a few orders to several orders of magnitude in the GW detection sensitivities over all frequency bands in the next hundred years\cite{ventitre}.

The gravitational deflection has already been applied to astrophysics and cosmology as gravitational lensing to weigh the lensing sources\cite{ventiquattro}. It becomes an important tool of astrophysics and cosmology. Besides GW observations, electromagnetic observations on black holes have been proposed (Sec. 2). It is well-known that satellite positioning systems need to incorporate GR corrections. With the clock reaching $10^{-18}$ and beyond, application to measure the Earth gravity and the altitude could be realized.
\section{Cosmic Polarization Rotation, Lense-Thirring Frame Dragging and Black Hole Shadow Observation}

\subsection{Summary on CPR}\label{aba:sec1}

A review about cosmic polarization rotation (CPR), i.e. a rotation of the polarization angle (PA) for radiation traveling over large distances across the universe, was presented by Sperello di Serego Alighieri\cite{sperello0}. CPR is very relevant for this Symposium, since it would be observed if there were the pseudoscalar field coupling to electromagnetism, which Ni\cite{Nio77} found as a unique counter-example to the conjecture that any consistent Lorentz-invariant theory of gravity obeying the weak equivalence principle (WEP) would also obey the Einstein equivalence principle (EEP). In fact, since general relativity (GR) is based on the EEP, our confidence on EEP and GR would be greatly increased if we could show that there is no CPR, because in this case the EEP would be tested to the same high accuracy of the WEP. The search for CPR is important also because it would tell us if and how one of the three elementary pieces of information (direction, energy, and PA), which photons carry to us about the universe, is changed while they travel.
Since 1990 CPR has been searched using the polarization of radio galaxies, both in the radio and in the ultraviolet, and, more recently, using the polarization of the cosmic microwave background (CMB).
The results of a recent review on CPR \cite{diS15} and of a few updates were presented. In summary, the results so far are consistent with a null CPR with upper limits of the order of $1^{\circ}$.
Two current problems in CPR searches  were discussed. The first involves the PA calibration at CMB frequencies, which is becoming the limiting factor, imposing a systematic error of about $1^{\circ}$, which is larger than the statistical errors of the best CMB polarization experiments. Improvements are expected from more precise measurements of the polarization angle of celestial sources at CMB frequencies and a calibration source on a satellite\cite{Kau16}.
The second problem results from the unfortunate choice of the CMB community for a PA convention which is opposite to the standard one, adopted by all astronomers for many decades and enforced by the International Astronomical Union\cite{iau74}. This is causing obvious confusion and misuderstanding, particularly for CPR, for which results use both conventions. A recommendation has been issued that all astronomers, including CMB polarimetrists, use the standard PA convention\cite{iau15}.
Concerning CPR tests, improvements are expected by better targeted high resolution radio polarization measurements of radio galaxies and quasars, by more accurate ultraviolet polarization measurements of radio galaxies with the coming generation of giant optical telescopes and by future CMB polarization measurements.
 An update\cite{pan} of CPR constraint was presented from the analysis of 
                           the recent
measurements of sub-degree B-mode polarization in the cosmic microwave background from 100 square degrees of SPTpol
data\cite{keisler}. The CPR fluctuation constraint from the joint
ACTpol-BICEP2-POLARBEAR polarization data is 23.7 mrad ($1.36^{\circ}$)\cite{mei}.
With the new SPTpol data included, the CPR fluctuation constraint is updated to 17 mrad ($1^{\circ}$) with the scalar to tensor
ratio $r = 0.05 {\pm} 0.1$.\cite{pan}
\subsection{The GINGER Project and the LenseThirring measurement}
GINGER (Gyroscopes IN General Relativity) \cite{angela}  is based on an array of ring-lasers and is aiming at measuring the LenseThirring effect at the level of $1\%$. Large frame ring-lasers are at present the most sensitive device to measure absolute angular rotation, and has been already demonstrated that they have a sensitivity very close to what is necessary to measure the Lense-Thirring effect\cite{PRD84}.
At present it is under discussion the real construction of GINGER. At the same time, experimental activity is in progress; the large frame prototype GINGERino\cite{otto} is investigating if the GranSasso underground laboratory is adeguate for an experiment as GINGER, and the prototype called GP2 has been installed in Pisa in order to develop a suitable control strategy to constrain the geometry of the ring-laser and guarantee the long term stability of the scale factor of the ring-laser. The long term stability of the scale factor of the ring-laser is the most challenging experimental problem of the research activity around GINGER.

\begin{figure}[pb]
\centerline{\psfig{file=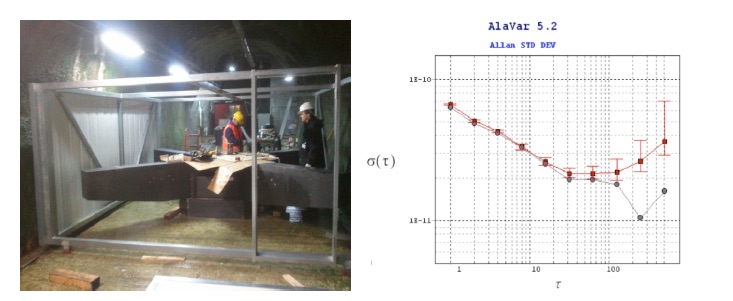,width=13.5cm}}
\vspace*{8pt}
\caption{ Left: The construction of GINGERino inside le underground GranSasso laboratory. Right: The preliminary Allan deviation, expressed in rad/s. \label{f1}}
\end{figure}

\subsection{The Greenland Telescope Project and BH Shadow Observation}
The size and shape of the shadow cast by a black hole event horizon is determined by the null geodesics and directly related to the background spacetime metric. Direct imaging of such shadow image and the test of physics
in strong gravity is one of the important goals in modern astronomy.  The ongoing Greenland Telescope (GLT)\cite{pu0} project in
Academia Sinica Institute of Astronomy and Astrophysics (ASIAA) is devoting itself to this exciting area \cite{pu1}. 
The target source of GLT project is the supermassive black hole located at the center of M87. 
The baseline of the future telescope in Greenland (hence the name Greenland Telescope), the Submillimeter Array in Hawaii, 
and the Atacama Large Millimeter/submillimeter Array in Chile, will play a key role for future Very Long Baseline Interferometry (VLBI)
observation. With longest baseline $>9000$ km, the angular resolution can reach $\sim20$
$\mu $as at 350 GHz, high enough to resolve the black hole shadow of M87, which has a estimated angular size $\sim40$
$\mu $as. 
The first light and related VLBI test of GLT in Thule, a northwest cost of Greenland, will be obtained in 2016. The GLT
will then be established at the Summit station in 2018/2019 \cite{pu5}. 
The GLT project is a collaborative project between ASIAA, Smithsonian Astrophysical Observatory, MIT Haystack Observatory, and National Radio Astronomy Observatory .

\begin{thebibliography}{00}
\bibitem{uno}See, e.g., W.-T. Ni, {\it{Int. J. Mod. Phys. D}} \textbf{25}, 1630003 (2016).
\bibitem{due} I. Shapiro, {\it{Phys. Rev. Lett.}}, \textbf{13} 789 (1964).
\bibitem{tre} B. Bertotti, L. Iess, and P. Tortora, {\it{Nature}} \textbf{425}, 374-376 (2003).
\bibitem{quattro} I. Ciufonlini, and E. C. Pavlis, {\it{Nature}} \textbf{431}, 958-960 (2004).
\bibitem{cinque} C. W. F. Everitt, et al., {\it{Phys. Rev. Lett.}} \textbf{106}, 221101 (2011). 
\bibitem{sei}W.-T. Ni, {\it{Phys. Rev. Lett.}}, \textbf{106}, 221101 (2011). 
\bibitem{sette}See, e.g., R. N. Manchester, {\it{Int. J. Mod. Phys. D}}, \textbf{24} (2015) 1530018.
\bibitem{otto}J. Belfi, et al., 'First Results of GINGERino', arXiv:1601.02874.
\bibitem{nove}W.-T. Ni, {\it{Phys.  Lett. A}}, 379, 1297 (2015); and references therein.
\bibitem{dieci}W.-T. Ni, {\it{Phys.  Lett. A}}, \textbf{378}, 3413 (2014).
\bibitem{undici}J.G. Williams, S. G. Turyshev, and D. H. Boggs, {\it{Int. J. Mod. Phys. D}}, \textbf{18}, 1129 (2009).
\bibitem{dodici}S. Schlamminger, et al., {\it{Phys. Rev. Lett.}}, \textbf{100}, 041101 (2008); and references therein.
\bibitem{tredici}http://smsc.cnes.fr/MICROSCOPE/index.htm.
\bibitem{quattordici}http://www.sstd.rl.ac.uk/fundphys/step/; http://einstein.stanford.edu/.
\bibitem{quindici}V. Rubin, W. K. Ford, Jr ,  {\it{Astrophys. J.}}, \textbf{159}, 379, (1970).
\bibitem{sedici}Rubin, V. C., Thonnard, N., and Ford, W. K., Jr., 1978, {\it{Astrophys. J.}}, \textbf{225}, L107.
\bibitem{diciassette}V. Rubin, N. Thonnard, W. K. Ford, Jr, (1980), {\it{Astrophys. J.}} \textbf{238}, 471.
\bibitem{diciotto}M. Milgrom, {\it{Astrophys. J.}}, \textbf{270} 365 (1983).
\bibitem{diciannove}See, e.g., M. Bucher and W.-T. Ni, {\it{Int. J. Mod. Phys. D}}, \textbf{24}, 1530030 (2015). 
\bibitem{venti}See, e.g., M. Davis, {\it{Int. J. Mod. Phys. D}}, \textbf{23}, 1430021 (2014).
\bibitem{ventuno}See, e.g., K. Sato and J. Yokoyama, {\it{Int. J. Mod. Phys. D}}, \textbf{24}, 1530025 (2015).
\bibitem{ventidue}BICEP2/Keck and Planck Collab., {\it{Phys. Rev. Lett.}}, \textbf{114}, 101301 (2015); C.-L. Kuo, this meeting.
\bibitem{wei0} W.-T. Ni, 'Sensitivities of gravitational-wave detection', \textit{these proceedings}.
\bibitem{GW}LIGO and Virgo Collaboration, {\em Phys. Rev. Lett.}, \textbf{116}, 6, 061102, 2016
\bibitem{ventitre}See, e.g., W.-T. Ni, {\it{Int. J. Mod. Phys. D}}, \textbf{24}, 1530031 (2015) (arXiv:1511.00231).
\bibitem{ventiquattro}See, e.g., T. Futamase, {\it{Int. J. Mod. Phys. D}}, \textbf{24}, 1530011 (2015).
\bibitem{sperello0} S. di Serego Alighieri, 'Gaining confidence on General Relativity with Cosmic Polarization Rotation', \textit{these proceedings}.
\bibitem{Nio77} Ni, W.-T., {\em Phys. Rev. Lett.} {\bf 38}, 301 (1977).
\bibitem{diS15} di Serego Alighieri, S., {\em IJMPD} {\bf 24}, 1530016 (2015).
\bibitem{Kau16} Kaufman, J.P., Keating, B.G., and Johnson, B.R., {\em MNRAS} {\bf 455}, 1981 (2016).

\bibitem{iau74} IAU Commission 40, {\em Polarization Definitions}, {\em Transactions of the IAU}, \textbf{Vol. XVB}, p. 166 (1974).
\bibitem{iau15} IAU Recom., http://www.iau.org/news/announcements/detail/ann16004 (2015).
\bibitem{pan}W.-P. Pan et al., 'New constraints on cosmic polarization rotation including SPTpol B-mode polarization observations', \textit{these proceedings}.
\bibitem{keisler} R. Keisler et al., \textit{Astrophysical Journal}, \textbf{807} (2015) 151
\bibitem{mei} H.-H. Mei et al., \textit{Astrophysical Journal}, \textbf{805} (2015) 107
\bibitem{angela} A. Di Virgilio, 'GINGER, An array of Ring Lasers to test General Relativity', \textit{these proceedings}
\bibitem{PRD84} F. Bosi et al.,{ \it{Phys. Rev. D}}, \textbf{84}, 12, 122002, 2011
\bibitem{pu0} H.-Y. Pu, 'Observing the Black Hole Shadow of M87 and the Greenland Telescope Project', \textit{these proceedings}.
\bibitem{pu1} ASIAA website for the GLT project:~url:http://vlbi.asiaa.sinica.edu.tw/project.php
\bibitem{pu5} Inoue M. et al. 2014 Greenland Telescope Project: Direct Confirmation of Black Hole with Sub-millimeter VLBI, RADIO SCIENCE: \textbf{49(7)}, 564-571, 2014-07\end{thebibliography}

\end{document}